\documentclass[twocolumn,aps,superscriptaddress,showpacs,floatfix]{revtex4-1}
\usepackage{graphicx}
\usepackage{dcolumn}
\usepackage{bm}
\usepackage[colorlinks,linkcolor=blue,anchorcolor=blue,citecolor=blue,urlcolor=blue]{hyperref}
\usepackage{amsmath}
\usepackage{multirow}
\begin{document}

\preprint{APS/123-QED}

\title{Systematic study of $\boldsymbol{\mathcal{\alpha}}$ decay of nuclei around $\boldsymbol{Z=82}$, $\boldsymbol{N=126}$ shell closure within the cluster-formation model and proximity potential 1977 formalism}

\author{Jun-Gang Deng}
\affiliation{School of Nuclear Science and Technology, University of South China, 421001 Hengyang, People's Republic of China}
\author{Jie-Cheng Zhao}
\affiliation{School of Nuclear Science and Technology, University of South China, 421001 Hengyang, People's Republic of China}
\author{Peng-Cheng Chu}
\email{kyois@126.com }
\affiliation{School of Science, Qingdao Technological University, 266000 Qingdao, People's Republic of China}
\author{Xiao-Hua Li}
\email{lixiaohuaphysics@126.com }
\affiliation{School of Nuclear Science and Technology, University of South China, 421001 Hengyang, People's Republic of China}
\affiliation{Cooperative Innovation Center for Nuclear Fuel Cycle Technology $\&$ Equipment, University of South China, 421001 Hengyang, People's Republic of China}
\affiliation{Key Laboratory of Low Dimensional Quantum Structures and Quantum Control, Hunan Normal University, 410081 Changsha, People's Republic of China}

\begin{abstract}
In the present work, we systematically study the $\mathcal{\alpha}$ decay preformation factors $P_{\alpha}$ within the cluster-formation model and $\mathcal{\alpha}$ decay half-lives by the proximity potential 1977 formalism for nuclei around $Z=82$, $N=126$ closed shells. The calculations show that the realistic $P_{\alpha}$ is linearly dependent on the product of valance protons (holes) and valance neutrons (holes) $N_pN_n$. It is consistent with our previous works [X.-D. Sun \textit{et al.}, \href{https://journals.aps.org/prc/abstract/10.1103/PhysRevC.94.024338}{ Phys. Rev. C 94, 024338 (2016)}, J.-G. Deng \textit{et al.}, \href{https://journals.aps.org/prc/abstract/10.1103/PhysRevC.96.024318}{Phys. Rev. C 96, 024318 (2017)}], which $P_{\alpha}$ are model-dependent and extracted from the ratios of calculated $\mathcal{\alpha}$ half-lives to experimental data.
Combining with our previous works, we confirm that the valance proton-neutron interaction plays a key role in the $\mathcal{\alpha}$ preformation for nuclei around  $Z=82$, $N=126$ shell closures whether the $P_{\alpha}$ is model-dependent or microcosmic. In addition, our calculated $\mathcal{\alpha}$ decay half-lives by using the proximity potential 1977 formalism taking $P_{\alpha}$ evaluated by the cluster-formation model can well reproduce the experimental data and significantly reduce the errors.
\end{abstract}

\maketitle

\section{Introduction}
In 1928, the phenomenon of $\mathcal{\alpha}$ decay for nuclei was independently explained by Gurney and Condon \cite{Gur28} and Gamow \cite{Gamow1928} using the quantum tunnel theory. Since then, $\mathcal{\alpha}$ decay has long been perceived as one of the most powerful tools to investigate unstable nuclei, neutron-deficient nuclei and superheavy nuclei, and has been an active area of research of nuclear physics \cite{PhysRevLett.112.172501,PhysRevLett.112.092501,PhysRevC.92.051301,Ni2015108,PhysRevC.84.064608, PhysRevC.85.044608,1674-1137-41-12-124109,PhysRevC.80.057301,Rom.Rep.Phys123,0954-3899-40-6-065105,0954-3899-42-7-075106,PhysRevC.93.044326,SALEHAHMED2017103,PhysRevC.95.034323,Guo2015110,PhysRevC.74.017304,PhysRevC.77.054318,BAO201485,PhysRevC.81.064309,PhysRevC.92.064301}. 

Within the Gamow\rq{}s theory, the $\mathcal{\alpha}$ decay process is  described as a preformed $\mathcal{\alpha}$ particle penetrating the Coulomb barrier. Thus an $\mathcal{\alpha}$ preformation factor should be introduced into $\mathcal{\alpha}$ decay theories, which denotes the probability of an $\mathcal{\alpha}$ cluster preformation. There are a lot of models devoted to determining $\mathcal{\alpha}$ preformation factors. Microscopically, $\mathcal{\alpha}$ preformation factors can be calculated by the overlap between initial wavefunction and $\mathcal{\alpha}$ decaying wavefunction \cite{ LOVAS1998265}. In the R-matrix method, the $\mathcal{\alpha}$ preformation can be obtained from the initial tailored wavefunction of the parent nucleus \cite{PhysRevLett.69.37,DODIGCRNKOVIC1985419,TONOZUKA197945,DODIGCRNKOVIC1989533,VARGA1992421}. R$\ddot{\rm{o}}$pke \textit{et al.} \cite{ PhysRevC.90.034304} and Xu \textit{et al.} \cite{PhysRevC.93.011306} calculated $\mathcal{\alpha}$ preformation factors using an approach of the Tohsaki-Horiuchi-Schuck-R$\ddot{\rm{o}}$pke wavefunction, which was also successfully used to describe the cluster structure of light nuclei. In the cluster model, the $\mathcal{\alpha}$ preformation factor is tread as a  constant less than one for a certain type of nuclei and the value of even-even nuclei$>$odd-$A$ nuclei$>$doubly-odd nuclei \cite{PhysRevC.45.2247,BUCK199353,PhysRevC.73.041301,PhysRevC.74.014304,PhysRevC.81.024315,PhysRevC.83.044317}. Xu and Ren systematically studied the $\mathcal{\alpha}$ decay of medium mass nuclei using the density-dependent cluster model (DDCM) \cite{XU2005303}. Their results indicated that the $\mathcal{\alpha}$ preformation factors are 0.43 for even-even nuclei, 0.35 for odd-$A$ nuclei, and 0.18 for doubly-odd nuclei. Because of the complicated structure of quantum many-body systems, phenomenologically, the $\mathcal{\alpha}$ preformation factors are extracted from the ratios of calculations to experimental $\mathcal{\alpha}$ decay half-lives \cite{PhysRevC.84.027303,Qian2013,PhysRevC.80.064325,PhysRevC.80.064325}. Nevertheless, the obtained preformation factors are strongly model-dependent.

Recently, Ahmed \textit{et al.} presented a new quantum-mechanical theory named cluster-formation model (CFM) to calculate the $\mathcal{\alpha}$ preformation factors $P_{\alpha}$ of even-even nuclei \cite{Rom.Rep.Phys123,0954-3899-40-6-065105}, which suggests that the initial state of the parent nucleus should be a linear combination of different possible clusterization states. They successfully determined the $P_{\alpha}= 0.22 $ for $^{212}$Po using CFM, which could well reproduce the calculations of Varga \textit{et al.} \cite{PhysRevLett.69.37,VARGA1992421}, and value of Ni and Ren  \cite{NI2009145} in different microscopic ways. Very recently, Ahmed \textit{et al.} and Deng \textit{et al.} extended CFM to odd-$A$ and doubly-odd nuclei through modifying the formation energy of interior $\mathcal{\alpha}$ cluster for various types of nuclei (i.e. even $Z$- odd $N$, odd $Z$-even $N$ and doubly-odd nuclei) and considering the effects of unpaired nucleon \cite{SALEHAHMED2017103,0954-3899-42-7-075106,PhysRevC.93.044326,Alsaif2016}. In 2011, Seif \textit{et al.} have put forward that the $\mathcal{\alpha}$ preformation factor is linearly dependent on $N_pN_n$ for even-even nuclei around proton $Z=82$, neutron $N=126$ closed shells, where $N_p$ and $N_n$ denote valance protons (holes) and valance neutrons (holes) \cite{PhysRevC.84.064608}. In our previous works, the extracted $\mathcal{\alpha}$ preformation factors from ratios of calculated $\mathcal{\alpha}$ decay half-life to experimental data for cases of odd-$A$ and doubly-odd nuclei $\mathcal{\alpha}$ decay also satisfy this relationship \cite{PhysRevC.94.024338,PhysRevC.96.024318}. It is interesting to validate whether the realistic $\mathcal{\alpha}$ preformation factor within CFM is also proportional to $N_pN_n$. In addition, many researchers adopted the Coulomb and proximity potential model (CPPM) to investigate $\mathcal{\alpha}$ decay leaving $P_{\alpha}$ out of consideration or assuming as $P_{\alpha}=1$, thus the deviations between calculated $\mathcal{\alpha}$ decay half-lives and experimental data were considerable \cite{PhysRevC.90.064603,Yao2015,PhysRevC.93.024612}. For confirming CFM and diminishing the difference between theoretical and experimental data, we also calculate $\mathcal{\alpha}$ decay half-lives within the Proximity potential 1977 formalism (Prox.1977) \cite{BLOCKI1977427} taking $P_{\alpha}=1$ and the realistic $P_{\alpha}$ evaluated by CFM, respectively. Our calculated $\mathcal{\alpha}$ decay half-lives within Prox.1977 taking $P_{\alpha}$ evaluated by CFM can  significantly reduce the deviations between calculations and experimental data.

This article is organized as follows. In next section, the theoretical framework of the CFM, $\mathcal{\alpha}$ decay half-life and Prox.1977 are briefly presented. The detailed calculations and discussion are given in Sec. \ref{section 3}. In this section, we investigate the $\mathcal{\alpha}$ preformation factors from the viewpoint of the valence proton-neutron interaction, and calculate $\mathcal{\alpha}$ decay half-lives by Prox.1977 with $P_{\alpha}=1$ and $P_{\alpha}$ calculated by CFM, respectively. Sec. \ref{section 4} is a brief summary.

\section{THEORETICAL FRAMEWORK}
\label{section 2}
\subsection{the cluster-formation model}
Within the cluster-formation model (CFM) \cite{Rom.Rep.Phys123,0954-3899-40-6-065105,0954-3899-42-7-075106,PhysRevC.93.044326,SALEHAHMED2017103}, the total clsuterization state $\Psi$ of parent nuclei is assumed as a linear combination of all its n possible clusterization states $\Psi_i$. It can be represented as
\begin{equation}
\
\Psi=\sum_{i=1}^n a_i{\Psi}_i
,\label{subeq:1}
\end{equation}
\begin{equation}
\
a_i={\int{\Psi}_i^*{\Psi} d\tau},
\label{subeq:1}
\end{equation}
where $a_i$ denotes the superposition coefficient of ${\Psi}_i$, on the basis of orthogonality condition,
\begin{equation}
\
\sum_{i=1}^n |{a}_i|^2=1
.\label{subeq:1}
\end{equation}
The total wavefunction is an eigenfunction of the total Hamiltonian $H$. Similarly, $H$ can be expressed as 
\begin{equation}
\
H=\sum_{i=1}^n H_i
,\label{subeq:1}
\end{equation}
where $H_i$ is the Hamiltonian for the ith clusterization state ${\Psi}_i$.
On account of the all clusterizations describing the same nucleus, they are assumed as sharing a same total energy $E$ of the total wavefunction. Thus the total energy $E$ can be expressed as
\begin{equation}
\
E=\sum_{i=1}^n|{a}_i|^2E= \sum_{i=1}^nE_{fi}
,\label{subeq:1}
\end{equation}
where $E_{fi}$ denotes the formation energy of cluster in the ith clusterization state ${\Psi}_i$. Therefore, the $\mathcal{\alpha}$ preformation factor can be defined as
\begin{equation}
\
P_{\alpha}=|{a}_{\alpha}|^2= \frac{E_{f\alpha}}{E}
,\label{subeq: 6}
\end{equation}
where $a_{\alpha}$ denotes the coefficient of the $\mathcal{\alpha}$ clusterization state. $E_{f\alpha}$ is the formation energy of the $\mathcal{\alpha}$ cluster. $E$ is composed of the $E_{f\alpha}$ and the interaction energy between $\mathcal{\alpha}$ cluster and daughter nuclei.
In the framework of CFM \cite{Rom.Rep.Phys123,0954-3899-40-6-065105,0954-3899-42-7-075106,PhysRevC.93.044326,SALEHAHMED2017103}, the $\mathcal{\alpha}$ cluster formation energy $E_{f\alpha}$ and total energy $E$ of considered system can be expressed as four different cases.

\begin{subequations}
\label{eq:whole}
Case I for even-even nuclei,
\begin{eqnarray}
E_{f\alpha}=&&3B(A,Z)+B(A-4,Z-2)\nonumber\\
&&-2B(A-1,Z-1)-2B(A-1,Z),\label{subeq:2}
\end{eqnarray}
\begin{equation}
E=B(A,Z)-B(A-4,Z-2),\label{subeq:1}
\end{equation}
case II for even $Z$-odd $N$ i.e. even-odd nuclei,
\begin{eqnarray}
E_{f\alpha}=&&3B(A-1,Z)+B(A-5,Z-2)\nonumber\\
&&-2B(A-2,Z-1)-2B(A-2,Z),\label{subeq:2}
\end{eqnarray}
\begin{equation}
E=B(A,Z)-B(A-5,Z-2),\label{subeq:1}
\end{equation}
case III for odd $Z$-even $N$ i.e. odd-even nuclei,
\begin{eqnarray}
E_{f\alpha}=&&3B(A-1,Z-1)+B(A-5,Z-3)\nonumber\\
&&-2B(A-2,Z-2)-2B(A-2,Z-1),\label{subeq:2}
\end{eqnarray}
\begin{equation}
E=B(A,Z)-B(A-5,Z-3),\label{subeq:1}
\end{equation}
case IV for doubly-odd nuclei,
\begin{eqnarray}
E_{f\alpha}=&&3B(A-2,Z-1)+B(A-6,Z-3)\nonumber\\
&&-2B(A-3,Z-2)-2B(A-3,Z-1),\label{subeq:2}
\end{eqnarray}
\begin{equation}
E=B(A,Z)-B(A-6,Z-3),\label{subeq:1}
\end{equation}
\end{subequations}
where $B(A, Z)$ denotes the binding energy of nucleus with the mass number $A$ and proton number $Z$.
\subsection{$\mathcal{\alpha}$ decay half-life and proximity potential 1977 formalism}
The $\mathcal{\alpha}$ decay half-life can be calculated by decay width $\Gamma$ or decay constant $\mathcal{\lambda}$ and expressed as
\begin{equation}
\
T_{1/2}=\frac{{\hbar}ln2}{\Gamma}=\frac{ln2}{\lambda}
,\label{subeq:1}
\end{equation}
where $\hbar$ is the Planck constant. In the framework of the Proximity potential 1977 formalism (Prox.1977) \cite{BLOCKI1977427}, the $\mathcal{\alpha}$ decay constant $\mathcal{\lambda}$ is calculated by
\begin{equation}
\
\lambda=P_{\alpha}{\nu}P
,\label{subeq:1}
\end{equation}
where $P_{\alpha}$ denotes $\mathcal{\alpha}$ preformation factors. In CPPM, the $P_{\alpha}$ is left out of consideration or assumed as $P_{\alpha}=1$. The assault frequency $\mathcal{\nu}$ can be obtained by the oscillation frequency $\mathcal{\omega}$ \cite{PhysRevC.81.064309}, and expressed as
\begin{equation}
\
\nu=\frac{\omega}{2\pi}=\frac{(2n_r+l+\frac{3}{2})\hbar}{2\pi\mu{R_n^2}}=\frac{(G+\frac{3}{2})\hbar}{1.2\pi\mu{R_0^2}}
,\label{subeq:1}
\end{equation}
where $\mu=\frac{{m_d}{m_{\alpha}}}{{m_d}+{m_{\alpha}}}$ denotes the reduced mass between daughter nucleus and preformed $\mathcal{\alpha}$ particle with the mass of daughter nucleus $m_d$ and $\mathcal{\alpha}$ particle $m_{\alpha}$. The nucleus root-mean-square (rms) radius $R_n=\sqrt{\frac{3}{5}}R_0$ with $R_0=1.240A^{1/3}(1+\frac{1.646}{A}-0.191\frac{A-2Z}{A})$ \cite{PhysRevC.62.044610}, where $A$ and $Z$ are mass number and proton number of parent nucleus. $G=2n_r+l$ denotes the principal quantum number with radial quantum number $n_r$ and angular momentum quantum number $l$. For $\mathcal{\alpha}$ decay \cite{PhysRevC.69.024614}, $G$ can be obtained by 

\begin{equation}
\
G=2n_r+l=\left\{\begin{array}{lll}
18,&$N$\leq82,\\
20,&82<$N$\leq126,\\
22,&$N$>126.
\end{array}\right.
\label{subeq:1}
\end {equation}
$P$, the semiclassical Wentzel-Kramers-Brillouin (WKB) barrier penetrate probability, can be calculated by
\begin{equation}
\
P=\exp(-2{\int_{r_{\text{in}}}^{r_{\text{out}}} k(r) dr})
,\label{subeq:1}
\end{equation}
where $k(r)=\sqrt{\frac{2\mu}{{\hbar}^2}|Q_{\alpha}-V(r)|}$ is the wave number of the $\mathcal{\alpha}$ particle. $r$ is the center of mass distance between the daughter nucleus and the preformed $\mathcal{\alpha}$ particle. $V(r)$ and $Q_{\alpha}$ are the total $\mathcal{\alpha}$-core potential and $\mathcal{\alpha}$ decay energy, respectively. $r_{\text{in}}$ and $r_{\text{out}}$ are the classical turning points, they satisfy the conditions $V (r_{\text{in}}) = V (r_{\text{out}}) =Q_{\alpha}$.

The total interaction potential $V(r)$ between $\mathcal{\alpha}$ particle and daughter nucleus is composed of three parts: the nuclear potential $V_N(r)$, the Coulomb potential $V_C(r)$ and the centrifugal potential $V_l(r)$. It can be expressed as
\begin{equation}
\
V(r)=V_N(r)+V_C(r)+V_l(r)
.\label{subeq:1}
\end {equation}

The Coulomb potential $V_C(r)$ is hypothesized as the potential of an uniformly charged sphere with sharp radius $R$ and expressed as
\begin{equation}
\
V_C(r)=\left\{\begin{array}{ll}

\frac{Z_dZ_{\alpha}e^2}{2R}[3-(\frac{r}{R})^2],&\text{{r}\textless{R}},\\

\frac{Z_dZ_{\alpha}e^2}{r},&\text{{r}\textgreater{R}},

\end{array}\right.
\label{subeq:1}
\end {equation}
where $R=R_1+R_2$ with $R_i=1.28A_i^{1/3}-0.76+0.8A_i^{-1/3}(i=1,2)$. $R_1$ and $R_2$ denote the radius of daughter nucleus and $\mathcal{\alpha}$ particle, respectively. $Z_d$ and $Z_{\alpha}$ are the proton number of daughter nucleus and $\mathcal{\alpha}$ particle, respectively.

Because $l(l+1){\to}(l+1/2)^2$ is a necessary corrections for one-dimensional problems \cite{1995JMP....36.5431M}, we adopt the Langer modified centrifugal barrier $V_l(r)$, which can be expressed as
\begin{equation}
\
V_l(r)=\frac{{\hbar}^2(l+1/2)^2}{2{\mu}r^2}
,\label{subeq:1}
\end {equation}
where $l$ is the angular momentum taken away by $\mathcal{\alpha}$ particle. On the basis of the conservation laws of angular momentum and parity \cite{PhysRevC.79.054614}, the minimum angular momentum $l_{\text{min}}$ taken away by the $\mathcal{\alpha}$ particle can be obtained by
\begin{equation}
\
l_{\text{min}}=\left\{\begin{array}{llll}

{\Delta}_j,&\text{for even${\Delta}_j$ and ${\pi}_p$= ${\pi}_d$},\\

{\Delta}_j+1,&\text{for even${\Delta}_j$ and ${\pi}_p$$\ne$${\pi}_d$},\\

{\Delta}_j,&\text{for odd${\Delta}_j$ and ${\pi}_p$$\ne$${\pi}_d$},\\

{\Delta}_j+1,&\text{for odd${\Delta}_j$ and ${\pi}_p$= ${\pi}_d$},

\end{array}\right.
\label{subeq:1}
\end {equation}
where ${\Delta}_j= |j_p-j_d|$, $j_p$, $\pi_p$, $j_d$, $\pi_d$ denote the spin and parity values of the parent and daughter nuclei, respectively.

The nuclear potential $V_N(r)$ is obtained by
\begin{equation}
\
V_N(r)=4\pi\gamma{b}\bar{R}\phi(\xi)
,\label{subeq:1}
\end {equation}
where $\gamma$, the surface energy coefficient, is obtained by the Myers and $\acute{\text{S}}$wiatecki formula \cite{1} and expressed as
\begin{equation}
\
\gamma={\gamma}_0(1-k_sI^2)
,\label{subeq:1}
\end {equation}
where $I$ denote the isospin of the parent nucleus. The surface energy constant ${\gamma}_0=0.9517$ Mev/${\text{fm}}^2$ and surface asymmetry constant $k_s=1.7826$ \cite{1}. The mean curvature radius $\bar{R}$ can be obtained by 
\begin{equation}
\
\bar{R}=\frac{C_1C_2}{C_1+C_2}
,\label{subeq:1}
\end {equation}
where $C_i=R_i[1-{(\frac{b}{R_i})}^2] (i=1,2)$ with $C_1$ and $C_2$ representing the matter radius of daughter nucleus and $\mathcal{\alpha}$ particle, respectively. $b$ is the diffuseness of nuclear surface, which is taken as unity. The universal function $\phi(\xi)$ is expressed as
\begin{equation}
\
\phi(\xi)=\left\{\begin{array}{ll}

-\frac{1}{2}{(\xi-2.54)}^2-0.0852{(\xi-2.54)}^3,&\xi\leq1.2511,\\

-3.437\exp(-\frac{\xi}{0.75}),&\xi\geq1.2511,

\end{array}\right.
\label{subeq:1}
\end {equation}
where $\xi=(r-C_1-C_2)/b$ denotes the minimum separation distance. 

\section{RESULTS AND DISCUSSION}
\label{section 3}

\begin{figure}[b]
\includegraphics[width=8.5cm]{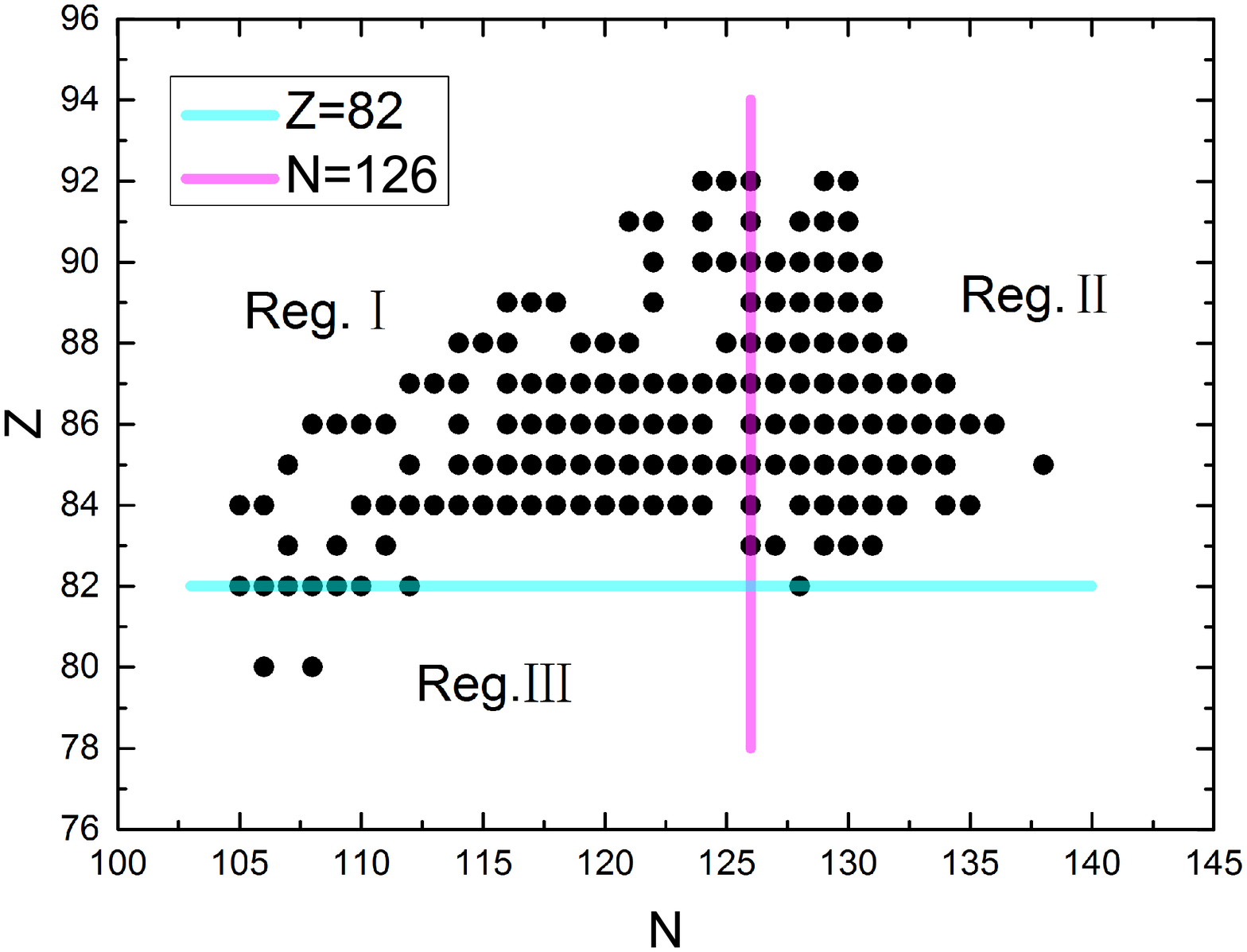}
\caption{(color online) Nuclide chart is divided into three regions. The cyan and magenta lines denote the $Z=82$, $N=126$ nuclear shell closures, respectively.}
\label{fig 1}
\end{figure}

\begin{figure}[b]
\includegraphics[width=8.5cm]{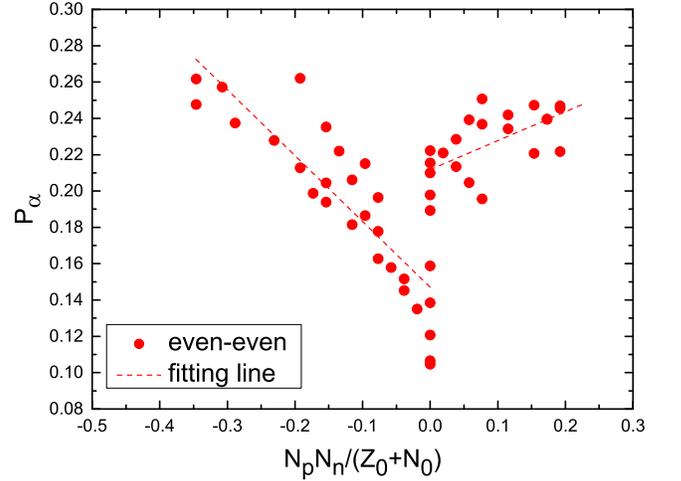}
\caption{(color online) The linear relationship between $\mathcal{\alpha}$ preformation factors and $\frac{N_p N_n}{Z_0+N_0}$. $N_p$ and $N_n$ represent valence protons (holes) and neutrons (holes) of parent nucleus, respectively. $Z_0$ and $N_0$ mean the magic numbers of proton and neutron, respectively. The dash lines represent the fittings of $\mathcal{\alpha}$ preformation factors.}
\label{fig 2}
\end{figure}

\begin{figure}[b]
\includegraphics[width=8.5cm]{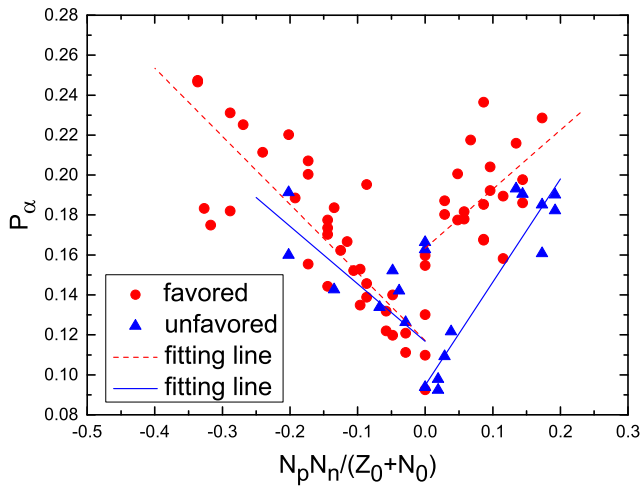}
\caption{(color online) Same as Fig. \ref{fig 2}, but it depicts linear relationships between $P_{\alpha}$ and $\frac{N_p N_n}{Z_0+N_0}$ of odd-$A$ nuclei. The red circle and blue triangle represent the cases of favored and unfavored $\mathcal{\alpha}$ decay, respectively. The red dash and blue solid lines represent the fittings of $\mathcal{\alpha}$ preformation factors for cases of favored and unfavored $\mathcal{\alpha}$ decay, respectively.}
\label{fig 3}
\end{figure}

\begin{figure}[b]
\includegraphics[width=8.5cm]{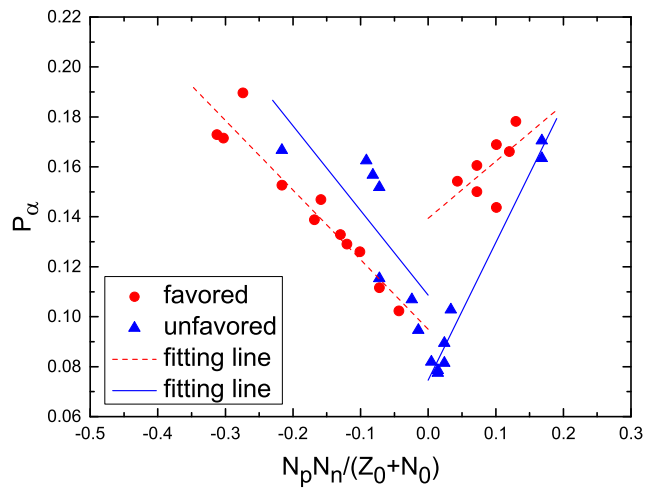}
\caption{(color online) Same as Fig. \ref{fig 2} and \ref{fig 3}, but it depicts linear relationships between $P_{\alpha}$ and $\frac{N_p N_n}{Z_0+N_0}$ of doubly-odd nuclei.}
\label{fig 4}
\end{figure}

\begin{table*}
\caption{Calculations of $\mathcal{\alpha}$ decay half-lives and the $\mathcal{\alpha}$ preformation factors of even-even nuclei in Region I, II and III around $Z=82$, $N=126$ closed shells. The experimental $\mathcal{\alpha}$ decay half-lives, spin and parity are taken from the latest evaluated nuclear properties table NUBASE2016 \cite{1674-1137-41-3-030001}, the $\mathcal{\alpha}$ decay energies are taken from the latest evaluated atomic mass table AME2016 \cite{1674-1137-41-3-030002,1674-1137-41-3-030003}. The $\mathcal{\alpha}$ preformation factors ${P_{\alpha}}$ are calculated within the CFM \cite{Rom.Rep.Phys123,0954-3899-40-6-065105,0954-3899-42-7-075106,PhysRevC.93.044326,SALEHAHMED2017103}.}
\label{table 1}
\begin{ruledtabular}
\begin{tabular}{ccccccccc}
{$\mathcal{\alpha}$ transition} & $Q_{\alpha}$ (MeV) & ${j^{\pi}_{p}}\to{j^{\pi}_{d}}$ &$l_{\text{min}}$ &${P_{\alpha}}$ & $T^{\text{expt}}_{1/2}$ (s)&${T_{1/2}^{\text{calc1}}}$ (s)& ${T_{1/2}^{\text{calc2}}}$ (s)& ${T_{1/2}^{\text{calc3}}}$ (s)\\
 \hline
 \noalign{\global\arrayrulewidth1pt}\noalign{\global\arrayrulewidth0.4pt} \multicolumn{9}{c}{\textbf{Nuclei in Region I}}\\
$^{190}$Po$\to^{186}$Pb$$&7.693&${0^+}\to{0^+}$ &0&0.262&$2.46\times10^{-3}$&$5.97\times10^{-4}$&$2.28\times10^{-3}$&$2.75\times10^{-3}$\\
$^{194}$Po$\to^{190}$Pb$$&6.987&${0^+}\to{0^+}$ &0&0.235&$3.92\times10^{-1}$&$1.31\times10^{-1}$&$5.56\times10^{-1}$&$6.45\times10^{-1}$\\
$^{196}$Po$\to^{192}$Pb$$&6.658&${0^+}\to{0^+}$ &0&0.222&$5.67\times10^{0}$&$2.19\times10^{0}$&$9.87\times10^{0}$&$1.12\times10^{1}$\\
$^{198}$Po$\to^{194}$Pb$$&6.310&${0^+}\to{0^+}$ &0&0.206&$1.85\times10^{2}$&$5.61\times10^{1}$&$2.72\times10^{2}$&$2.97\times10^{2}$\\
$^{200}$Po$\to^{196}$Pb$$&5.981&${0^+}\to{0^+}$ &0&0.187&$6.20\times10^{3}$&$1.57\times10^{3}$&$8.44\times10^{3}$&$8.66\times10^{3}$\\
$^{202}$Po$\to^{198}$Pb$$&5.700&${0^+}\to{0^+}$ &0&0.178&$1.39\times10^{5}$&$3.42\times10^{4}$&$1.92\times10^{5}$&$1.95\times10^{5}$\\
$^{204}$Po$\to^{200}$Pb$$&5.485&${0^+}\to{0^+}$ &0&0.158&$1.88\times10^{6}$&$4.18\times10^{5}$&$2.64\times10^{6}$&$2.49\times10^{6}$\\
$^{206}$Po$\to^{202}$Pb$$&5.327&${0^+}\to{0^+}$ &0&0.145&$1.39\times10^{7}$&$2.85\times10^{6}$&$1.96\times10^{7}$&$1.77\times10^{7}$\\
$^{208}$Po$\to^{204}$Pb$$&5.216&${0^+}\to{0^+}$ &0&0.135&$9.15\times10^{7}$&$1.15\times10^{7}$&$8.51\times10^{7}$&$7.47\times10^{7}$\\
$^{194}$Rn$\to^{190}$Po$$&7.862&${0^+}\to{0^+}$ &0&0.262&$7.80\times10^{-4}$&$1.04\times10^{-3}$&$3.99\times10^{-3}$&$3.83\times10^{-3}$\\
$^{196}$Rn$\to^{192}$Po$$&7.617&${0^+}\to{0^+}$ &0&0.257&$4.70\times10^{-3}$&$5.89\times10^{-3}$&$2.29\times10^{-2}$&$2.28\times10^{-2}$\\
$^{200}$Rn$\to^{196}$Po$$&7.043&${0^+}\to{0^+}$ &0&0.228&$1.17\times10^{0}$&$5.19\times10^{-1}$&$2.28\times10^{0}$&$2.25\times10^{0}$\\
$^{202}$Rn$\to^{198}$Po$$&6.773&${0^+}\to{0^+}$ &0&0.213&$1.23\times10^{1}$&$5.26\times10^{0}$&$2.47\times10^{1}$&$2.43\times10^{1}$\\
$^{204}$Rn$\to^{200}$Po$$&6.547&${0^+}\to{0^+}$ &0&0.194&$1.03\times10^{2}$&$4.05\times10^{1}$&$2.09\times10^{2}$&$2.00\times10^{2}$\\
$^{206}$Rn$\to^{202}$Po$$&6.384&${0^+}\to{0^+}$ &0&0.181&$5.46\times10^{2}$&$1.86\times10^{2}$&$1.02\times10^{3}$&$9.84\times10^{2}$\\
$^{208}$Rn$\to^{204}$Po$$&6.260&${0^+}\to{0^+}$ &0&0.163&$2.33\times10^{3}$&$6.07\times10^{2}$&$3.73\times10^{3}$&$3.47\times10^{3}$\\
$^{210}$Rn$\to^{206}$Po$$&6.159&${0^+}\to{0^+}$ &0&0.152&$8.99\times10^{3}$&$1.62\times10^{3}$&$1.07\times10^{4}$&$1.00\times10^{4}$\\
$^{212}$Rn$\to^{208}$Po$$&6.385&${0^+}\to{0^+}$ &0&0.121&$1.43\times10^{3}$&$1.44\times10^{2}$&$1.19\times10^{3}$&$9.79\times10^{2}$\\
$^{202}$Ra$\to^{198}$Rn$$&7.880&${0^+}\to{0^+}$ &0&0.248&$4.10\times10^{-3}$&$4.50\times10^{-3}$&$1.82\times10^{-2}$&$1.65\times10^{-2}$\\
$^{204}$Ra$\to^{200}$Rn$$&7.637&${0^+}\to{0^+}$ &0&0.237&$6.00\times10^{-2}$&$2.62\times10^{-2}$&$1.10\times10^{-1}$&$1.04\times10^{-1}$\\
$^{208}$Ra$\to^{204}$Rn$$&7.273&${0^+}\to{0^+}$ &0&0.199&$1.27\times10^{0}$&$4.20\times10^{-1}$&$2.11\times10^{0}$&$2.00\times10^{0}$\\
$^{214}$Ra$\to^{210}$Rn$$&7.273&${0^+}\to{0^+}$ &0&0.139&$2.44\times10^{0}$&$3.25\times10^{-1}$&$2.34\times10^{0}$&$2.21\times10^{0}$\\
$^{212}$Th$\to^{208}$Ra$$&7.958&${0^+}\to{0^+}$ &0&0.205&$3.17\times10^{-2}$&$1.14\times10^{-2}$&$5.59\times10^{-2}$&$5.64\times10^{-2}$\\
$^{214}$Th$\to^{210}$Ra$$&7.827&${0^+}\to{0^+}$ &0&0.196&$8.70\times10^{-2}$&$2.84\times10^{-2}$&$1.45\times10^{-1}$&$1.63\times10^{-1}$\\
$^{216}$U$\to^{212}$Th$$&8.530&${0^+}\to{0^+}$ &0&0.215&$6.90\times10^{3}$&$1.06\times10^{-3}$&$4.93\times10^{-3}$&$5.83\times10^{-3}$\\
 \noalign{\global\arrayrulewidth1pt}\noalign{\global\arrayrulewidth0.4pt} \multicolumn{9}{c}{\textbf{Nuclei in Region II and III}}\\
$^{186}$Hg$\to^{182}$Pt$$&5.204&${0^+}\to{0^+}$ &0&0.247&$5.02\times10^{5}$&$1.86\times10^{5}$&$7.53\times10^{5}$&$7.67\times10^{5}$\\
$^{188}$Hg$\to^{184}$Pt$$&4.707&${0^+}\to{0^+}$ &0&0.239&$3.33\times10^{9}$&$1.56\times10^{8}$&$6.53\times10^{8}$&$6.53\times10^{8}$\\
$^{188}$Pb$\to^{184}$Hg$$&6.109&${0^+}\to{0^+}$ &0&0.222&$2.68\times10^{2}$&$6.70\times10^{1}$&$3.01\times10^{2}$&$3.16\times10^{2}$\\
$^{190}$Pb$\to^{186}$Hg$$&5.697&${0^+}\to{0^+}$ &0&0.215&$1.76\times10^{4}$&$5.17\times10^{3}$&$2.40\times10^{4}$&$2.44\times10^{4}$\\
$^{192}$Pb$\to^{188}$Hg$$&5.221&${0^+}\to{0^+}$ &0&0.210&$3.52\times10^{6}$&$1.54\times10^{6}$&$7.35\times10^{6}$&$7.29\times10^{6}$\\
$^{194}$Pb$\to^{190}$Hg$$&4.738&${0^+}\to{0^+}$ &0&0.198&$1.71\times10^{10}$&$1.23\times10^{9}$&$6.19\times10^{9}$&$5.79\times10^{9}$\\
$^{210}$Pb$\to^{206}$Hg$$&3.793&${0^+}\to{0^+}$ &0&0.107&$9.26\times10^{16}$&$1.20\times10^{16}$&$1.13\times10^{17}$&$5.67\times10^{16}$\\
$^{210}$Po$\to^{206}$Pb$$&5.408&${0^+}\to{0^+}$ &0&0.105&$1.20\times10^{7}$&$8.70\times10^{5}$&$8.31\times10^{6}$&$4.11\times10^{6}$\\
$^{212}$Po$\to^{208}$Pb$$&8.954&${0^+}\to{0^+}$ &0&0.221&$2.95\times10^{-7}$&$5.78\times10^{-8}$&$2.62\times10^{-7}$&$2.69\times10^{-7}$\\
$^{214}$Po$\to^{210}$Pb$$&7.834&${0^+}\to{0^+}$ &0&0.213&$1.64\times10^{-4}$&$7.03\times10^{-5}$&$3.30\times10^{-4}$&$3.23\times10^{-4}$\\
$^{216}$Po$\to^{212}$Pb$$&6.907&${0^+}\to{0^+}$ &0&0.205&$1.45\times10^{-1}$&$9.63\times10^{-2}$&$4.71\times10^{-1}$&$4.36\times10^{-1}$\\
$^{218}$Po$\to^{214}$Pb$$&6.115&${0^+}\to{0^+}$ &0&0.196&$1.86\times10^{2}$&$1.72\times10^{2}$&$8.77\times10^{2}$&$7.66\times10^{2}$\\
$^{214}$Rn$\to^{210}$Po$$&9.208&${0^+}\to{0^+}$ &0&0.228&$2.70\times10^{-7}$&$6.95\times10^{-8}$&$3.04\times10^{-7}$&$3.19\times10^{-7}$\\
$^{216}$Rn$\to^{212}$Po$$&8.198&${0^+}\to{0^+}$ &0&0.237&$4.50\times10^{-5}$&$3.42\times10^{-5}$&$1.44\times10^{-4}$&$1.53\times10^{-4}$\\
$^{218}$Rn$\to^{214}$Po$$&7.263&${0^+}\to{0^+}$ &0&0.234&$3.38\times10^{-2}$&$3.52\times10^{-2}$&$1.50\times10^{-1}$&$1.53\times10^{-1}$\\
$^{220}$Rn$\to^{216}$Po$$&6.405&${0^+}\to{0^+}$ &0&0.221&$5.56\times10^{1}$&$7.97\times10^{1}$&$3.61\times10^{2}$&$3.37\times10^{2}$\\
$^{222}$Rn$\to^{218}$Po$$&5.591&${0^+}\to{0^+}$ &0&0.222&$3.30\times10^{5}$&$6.49\times10^{5}$&$2.93\times10^{6}$&$2.68\times10^{6}$\\
$^{216}$Ra$\to^{212}$Rn$$&9.526&${0^+}\to{0^+}$ &0&0.239&$1.82\times10^{-7}$&$5.88\times10^{-8}$&$2.46\times10^{-7}$&$2.66\times10^{-7}$\\
$^{218}$Ra$\to^{214}$Rn$$&8.546&${0^+}\to{0^+}$ &0&0.242&$2.52\times10^{-5}$&$1.96\times10^{-5}$&$8.09\times10^{-5}$&$8.50\times10^{-5}$\\
$^{220}$Ra$\to^{216}$Rn$$&7.592&${0^+}\to{0^+}$ &0&0.240&$1.79\times10^{-2}$&$1.73\times10^{-2}$&$7.21\times10^{-2}$&$7.22\times10^{-2}$\\
$^{216}$Th$\to^{212}$Ra$$&8.072&${0^+}\to{0^+}$ &0&0.159&$2.60\times10^{-2}$&$4.11\times10^{-3}$&$2.59\times10^{-2}$&$1.94\times10^{-2}$\\
$^{218}$Th$\to^{214}$Ra$$&9.849&${0^+}\to{0^+}$ &0&0.251&$1.17\times10^{-7}$&$4.92\times10^{-8}$&$1.96\times10^{-7}$&$2.20\times10^{-7}$\\
$^{220}$Th$\to^{216}$Ra$$&8.953&${0^+}\to{0^+}$ &0&0.247&$9.70\times10^{-6}$&$8.11\times10^{-6}$&$3.28\times10^{-5}$&$3.43\times10^{-5}$\\
$^{218}$U$\to^{214}$Th$$&8.775&${0^+}\to{0^+}$ &0&0.189&$5.50\times10^{-4}$&$1.85\times10^{-4}$&$9.75\times10^{-4}$&$8.72\times10^{-4}$\\
$^{222}$U$\to^{218}$Th$$&9.478&${0^+}\to{0^+}$ &0&0.246&$4.70\times10^{-6}$&$1.79\times10^{-6}$&$7.30\times10^{-6}$&$7.40\times10^{-6}$\\

\end{tabular}
\end{ruledtabular}
\end{table*}

\begin{table*}
\caption{Same as Table \ref{table 1}, but for favored $\mathcal{\alpha}$ decay of odd-$A$ nuclei. `()' means uncertain spin and/or parity, `\#' means values estimated from trends in neighboring nuclides with the same $Z$ and $N$ parities, which are taken from NUBASE2016 \cite{1674-1137-41-3-030001}.}
\label{table 2}
\begin{ruledtabular}
\begin{tabular}{ccccccccc}
{$\mathcal{\alpha}$ transition} & $Q_{\alpha}$ (MeV) & ${j^{\pi}_{p}}\to{j^{\pi}_{d}}$ &$l_{\text{min}}$ &${P_{\alpha}}$ & $T^{\text{expt}}_{1/2}$ (s)&${T_{1/2}^{\text{calc1}}}$ (s)& ${T_{1/2}^{\text{calc2}}}$ (s)& ${T_{1/2}^{\text{calc3}}}$ (s)\\
\hline
\noalign{\global\arrayrulewidth1pt}\noalign{\global\arrayrulewidth0.4pt} \multicolumn{9}{c}{\textbf{Nuclei in Region I}}\\
$^{195}$Po$\to^{191}$Pb$$&6.745&${(3/2^-)}\to{(3/2^-)}$ &0&0.170&$4.92\times10^{0}$&$1.04\times10^{0}$&$6.11\times10^{0}$&$6.26\times10^{0}$\\
$^{197}$Po$\to^{193}$Pb$$&6.405&${(3/2^-)}\to{(3/2^-)}$ &0&0.162&$1.20\times10^{2}$&$2.30\times10^{1}$&$1.42\times10^{2}$&$1.44\times10^{2}$\\
$^{199}$Po$\to^{195}$Pb$$&6.075&${(3/2^-)}\to{3/2^-}$ &0&0.152&$4.36\times10^{3}$&$6.01\times10^{2}$&$3.95\times10^{3}$&$3.92\times10^{3}$\\
$^{201}$Po$\to^{197}$Pb$$&5.799&${3/2^-}\to{3/2^-}$ &0&0.139&$8.26\times10^{4}$&$1.14\times10^{4}$&$8.20\times10^{4}$&$7.77\times10^{4}$\\
$^{205}$Po$\to^{201}$Pb$$&5.325&${5/2^-}\to{5/2^-}$ &0&0.120&$1.53\times10^{7}$&$3.05\times10^{6}$&$2.55\times10^{7}$&$2.28\times10^{7}$\\
$^{207}$Po$\to^{203}$Pb$$&5.216&${5/2^-}\to{5/2^-}$ &0&0.111&$9.85\times10^{7}$&$1.18\times10^{7}$&$1.07\times10^{8}$&$9.33\times10^{7}$\\
$^{197}$At$\to^{193}$Bi$$&7.105&${(9/2^-)}\to{(9/2^-)}$ &0&0.220&$4.04\times10^{-1}$&$1.21\times10^{-1}$&$5.50\times10^{-1}$&$6.51\times10^{-1}$\\
$^{199}$At$\to^{195}$Bi$$&6.778&${9/2(^-)}\to{9/2(^-)}$ &0&0.200&$7.83\times10^{0}$&$1.92\times10^{0}$&$9.58\times10^{0}$&$1.09\times10^{1}$\\
$^{201}$At$\to^{197}$Bi$$&6.473&${(9/2^-)}\to{(9/2^-)}$ &0&0.177&$1.19\times10^{2}$&$3.07\times10^{1}$&$1.73\times10^{2}$&$1.85\times10^{2}$\\
$^{203}$At$\to^{199}$Bi$$&6.210&${9/2^-}\to{9/2^-}$ &0&0.167&$1.42\times10^{3}$&$3.95\times10^{2}$&$2.37\times10^{3}$&$2.53\times10^{3}$\\
$^{205}$At$\to^{201}$Bi$$&6.019&${9/2^-}\to{9/2^-}$ &0&0.146&$1.99\times10^{4}$&$2.76\times10^{3}$&$1.90\times10^{4}$&$1.88\times10^{4}$\\
$^{207}$At$\to^{203}$Bi$$&5.873&${9/2^-}\to{9/2^-}$ &0&0.132&$6.52\times10^{4}$&$1.28\times10^{4}$&$9.73\times10^{4}$&$9.38\times10^{4}$\\
$^{209}$At$\to^{205}$Bi$$&5.757&${9/2^-}\to{9/2^-}$ &0&0.121&$4.70\times10^{5}$&$4.49\times10^{4}$&$3.71\times10^{5}$&$3.53\times10^{5}$\\
$^{211}$At$\to^{207}$Bi$$&5.983&${9/2^-}\to{9/2^-}$ &0&0.093&$6.21\times10^{4}$&$3.23\times10^{3}$&$3.49\times10^{4}$&$2.76\times10^{4}$\\
$^{195}$Rn$\to^{191}$Po$$&7.694&${3/2^-}\to{(3/2^-)}$ &0&0.183&$7.00\times10^{-3}$&$3.45\times10^{-3}$&$1.88\times10^{-2}$&$1.51\times10^{-2}$\\
$^{197}$Rn$\to^{193}$Po$$&7.410&${(3/2^-)}\to{(3/2^-)}$ &0&0.182&$5.40\times10^{-2}$&$2.83\times10^{-2}$&$1.56\times10^{-1}$&$1.31\times10^{-1}$\\
$^{203}$Rn$\to^{199}$Po$$&6.629&${3/2^-\#}\to{(3/2^-)}$ &0&0.155&$6.58\times10^{1}$&$1.93\times10^{1}$&$1.24\times10^{2}$&$1.10\times10^{2}$\\
$^{207}$Rn$\to^{203}$Po$$&6.251&${5/2^-}\to{5/2^-}$ &0&0.135&$2.61\times10^{3}$&$6.95\times10^{2}$&$5.15\times10^{3}$&$4.64\times10^{3}$\\
$^{209}$Rn$\to^{205}$Po$$&6.155&${5/2^-}\to{5/2^-}$ &0&0.122&$1.00\times10^{4}$&$1.76\times10^{3}$&$1.44\times10^{4}$&$1.28\times10^{4}$\\
$^{199}$Fr$\to^{195}$At$$&7.816&${1/2^+\#}\to{1/2^+}$ &0&0.247&$6.60\times10^{-3}$&$3.09\times10^{-3}$&$1.25\times10^{-2}$&$1.33\times10^{-2}$\\
$^{201}$Fr$\to^{197}$At$$&7.519&${(9/2^-)}\to{(9/2^-)}$ &0&0.231&$6.28\times10^{-2}$&$2.75\times10^{-2}$&$1.19\times10^{-1}$&$1.28\times10^{-1}$\\
$^{203}$Fr$\to^{199}$At$$&7.274&${9/2^-}\to{9/2(^-)}$ &0&0.211&$5.50\times10^{-1}$&$1.83\times10^{-1}$&$8.67\times10^{-1}$&$9.20\times10^{-1}$\\
$^{205}$Fr$\to^{201}$At$$&7.054&${9/2^-}\to{(9/2^-)}$ &0&0.188&$3.82\times10^{0}$&$1.09\times10^{0}$&$5.80\times10^{0}$&$5.99\times10^{0}$\\
$^{207}$Fr$\to^{203}$At$$&6.894&${9/2^-}\to{9/2^-}$ &0&0.174&$1.55\times10^{1}$&$4.16\times10^{0}$&$2.39\times10^{1}$&$2.50\times10^{1}$\\
$^{209}$Fr$\to^{205}$At$$&6.777&${9/2^-}\to{9/2^-}$ &0&0.153&$5.66\times10^{1}$&$1.12\times10^{1}$&$7.29\times10^{1}$&$7.44\times10^{1}$\\
$^{211}$Fr$\to^{207}$At$$&6.662&${9/2^-}\to{9/2^-}$ &0&0.140&$2.13\times10^{2}$&$3.03\times10^{1}$&$2.16\times10^{2}$&$2.27\times10^{2}$\\
$^{213}$Fr$\to^{209}$At$$&6.905&${9/2^-}\to{9/2^-}$ &0&0.110&$3.43\times10^{1}$&$2.92\times10^{0}$&$2.66\times10^{1}$&$2.49\times10^{1}$\\
$^{203}$Ra$\to^{199}$Rn$$&7.735&${(3/2^-)}\to{(3/2^-)}$ &0&0.175&$3.60\times10^{-2}$&$1.28\times10^{-2}$&$7.34\times10^{-2}$&$5.70\times10^{-2}$\\
$^{209}$Ra$\to^{205}$Rn$$&7.143&${5/2^-}\to{5/2^-}$ &0&0.144&$4.71\times10^{0}$&$1.22\times10^{0}$&$8.49\times10^{0}$&$7.36\times10^{0}$\\
$^{205}$Ac$\to^{201}$Fr$$&8.096&${9/2^-\#}\to{(9/2^-)}$ &0&0.247&$8.00\times10^{-2}$&$2.14\times10^{-3}$&$8.67\times10^{-3}$&$9.22\times10^{-3}$\\
$^{207}$Ac$\to^{203}$Fr$$&7.849&${9/2^-\#}\to{9/2^-}$ &0&0.225&$3.10\times10^{-2}$&$1.21\times10^{-2}$&$5.39\times10^{-2}$&$5.81\times10^{-2}$\\
$^{211}$Ac$\to^{207}$Fr$$&7.619&${9/2^-}\to{9/2^-}$ &0&0.184&$2.13\times10^{-1}$&$6.05\times10^{-2}$&$3.30\times10^{-1}$&$3.71\times10^{-1}$\\
$^{213}$Pa$\to^{209}$Ac$$&8.395&${9/2^-\#}\to{(9/2^-)}$ &0&0.207&$7.00\times10^{-3}$&$1.21\times10^{-3}$&$5.82\times10^{-3}$&$6.84\times10^{-3}$\\
$^{215}$Pa$\to^{211}$Ac$$&8.235&${9/2^-\#}\to{9/2^-}$ &0&0.195&$1.40\times10^{-2}$&$3.44\times10^{-3}$&$1.76\times10^{-2}$&$2.35\times10^{-2}$\\
 \noalign{\global\arrayrulewidth1pt}\noalign{\global\arrayrulewidth0.4pt} \multicolumn{9}{c}{\textbf{Nuclei in Region II and III}}\\
$^{191}$Pb$\to^{187}$Hg$$&5.463&${(3/2^-)}\to{3/2(^-)}$ &0&0.160&$1.55\times10^{4}$&$7.72\times10^{4}$&$4.83\times10^{5}$&$4.73\times10^{5}$\\
$^{213}$Po$\to^{209}$Pb$$&8.536&${9/2^+}\to{9/2^+}$ &0&0.180&$3.71\times10^{-6}$&$6.82\times10^{-7}$&$3.78\times10^{-6}$&$3.97\times10^{-6}$\\
$^{215}$Po$\to^{211}$Pb$$&7.527&${9/2^+}\to{9/2^+}$ &0&0.177&$1.78\times10^{-3}$&$6.50\times10^{-4}$&$3.66\times10^{-3}$&$3.66\times10^{-3}$\\
$^{219}$Po$\to^{215}$Pb$$&5.916&${9/2^+\#}\to{9/2^+\#}$ &0&0.167&$2.19\times10^{3}$&$1.41\times10^{3}$&$8.43\times10^{3}$&$7.47\times10^{3}$\\
$^{213}$At$\to^{209}$Bi$$&9.254&${9/2^-}\to{9/2^-}$ &0&0.187&$1.25\times10^{-7}$&$2.40\times10^{-8}$&$1.28\times10^{-7}$&$1.40\times10^{-7}$\\
$^{215}$At$\to^{211}$Bi$$&8.178&${9/2^-}\to{9/2^-}$ &0&0.178&$1.00\times10^{-4}$&$1.60\times10^{-5}$&$8.98\times10^{-5}$&$8.85\times10^{-5}$\\
$^{217}$At$\to^{213}$Bi$$&7.202&${9/2^-}\to{9/2^-}$ &0&0.168&$3.26\times10^{-2}$&$2.15\times10^{-2}$&$1.28\times10^{-1}$&$1.14\times10^{-1}$\\
$^{219}$At$\to^{215}$Bi$$&6.342&${(9/2^-)}\to{(9/2^-)}$ &0&0.158&$5.98\times10^{1}$&$5.02\times10^{1}$&$3.17\times10^{2}$&$2.54\times10^{2}$\\
$^{215}$Rn$\to^{211}$Po$$&8.839&${9/2^+}\to{9/2^+}$ &0&0.182&$2.30\times10^{-6}$&$5.80\times10^{-7}$&$3.20\times10^{-6}$&$3.22\times10^{-6}$\\
$^{217}$Rn$\to^{213}$Po$$&7.888&${9/2^+}\to{9/2^+}$ &0&0.192&$5.40\times10^{-4}$&$2.90\times10^{-4}$&$1.51\times10^{-3}$&$1.51\times10^{-3}$\\
$^{215}$Fr$\to^{211}$At$$&9.541&${9/2^-}\to{9/2^-}$ &0&0.201&$8.60\times10^{-8}$&$2.46\times10^{-8}$&$1.23\times10^{-7}$&$1.38\times10^{-7}$\\
$^{217}$Fr$\to^{213}$At$$&8.470&${9/2^-}\to{9/2^-}$ &0&0.204&$1.68\times10^{-5}$&$1.34\times10^{-5}$&$6.58\times10^{-5}$&$7.00\times10^{-5}$\\
$^{219}$Fr$\to^{215}$At$$&7.449&${9/2^-}\to{9/2^-}$ &0&0.198&$2.00\times10^{-2}$&$2.05\times10^{-2}$&$1.04\times10^{-1}$&$9.96\times10^{-2}$\\
$^{217}$Ra$\to^{213}$Rn$$&9.161&${(9/2^+)}\to{9/2^+\#}$ &0&0.185&$1.63\times10^{-6}$&$4.50\times10^{-7}$&$2.43\times10^{-6}$&$2.38\times10^{-6}$\\
$^{215}$Ac$\to^{211}$Fr$$&7.746&${9/2^-}\to{9/2^-}$ &0&0.130&$1.70\times10^{-1}$&$1.88\times10^{-2}$&$1.44\times10^{-1}$&$1.15\times10^{-1}$\\
$^{217}$Ac$\to^{213}$Fr$$&9.832&${9/2^-}\to{9/2^-}$ &0&0.218&$6.90\times10^{-8}$&$2.47\times10^{-8}$&$1.14\times10^{-7}$&$1.35\times10^{-7}$\\
$^{219}$Ac$\to^{215}$Fr$$&8.827&${9/2^-}\to{9/2^-}$ &0&0.216&$1.18\times10^{-5}$&$7.63\times10^{-6}$&$3.53\times10^{-5}$&$3.75\times10^{-5}$\\
$^{219}$Th$\to^{215}$Ra$$&9.511&${9/2^+\#}\to{9/2^+\#}$ &0&0.189&$1.02\times10^{-6}$&$3.05\times10^{-7}$&$1.61\times10^{-6}$&$1.54\times10^{-6}$\\
$^{217}$Pa$\to^{213}$Ac$$&8.488&${9/2^-\#}\to{9/2^-\#}$ &0&0.155&$3.48\times10^{-3}$&$5.29\times10^{-4}$&$3.42\times10^{-3}$&$3.24\times10^{-3}$\\
$^{219}$Pa$\to^{215}$Ac$$&10.084&${9/2^-}\to{9/2^-}$ &0&0.236&$5.30\times10^{-8}$&$3.08\times10^{-8}$&$1.30\times10^{-7}$&$1.63\times10^{-7}$\\
$^{221}$Pa$\to^{217}$Ac$$&9.251&${9/2^-}\to{9/2^-}$ &0&0.229&$5.90\times10^{-6}$&$3.02\times10^{-6}$&$1.32\times10^{-5}$&$1.41\times10^{-5}$\\
$^{221}$U$\to^{217}$Th$$&9.889&${(9/2^+)}\to{9/2^+\#}$ &0&0.186&$6.60\times10^{-7}$&$1.83\times10^{-7}$&$9.82\times10^{-7}$&$8.87\times10^{-7}$\\

\end{tabular}
\end{ruledtabular}
\end{table*}

\begin{table*}
\caption{Same as Table \ref{table 1} and \ref{table 2}, but for unfavored $\mathcal{\alpha}$ decay of odd-$A$ nuclei.}
\label{table 3}
\begin{ruledtabular}
\begin{tabular}{ccccccccc}

{$\mathcal{\alpha}$ transition} & $Q_{\alpha}$ (MeV) & ${j^{\pi}_{p}}\to{j^{\pi}_{d}}$ &$l_{\text{min}}$ &${P_{\alpha}}$ & $T^{\text{expt}}_{1/2}$ (s)&${T_{1/2}^{\text{calc1}}}$ (s)& ${T_{1/2}^{\text{calc2}}}$ (s)& ${T_{1/2}^{\text{calc3}}}$ (s)\\
\hline
 \noalign{\global\arrayrulewidth1pt}\noalign{\global\arrayrulewidth0.4pt} \multicolumn{9}{c}{\textbf{Nuclei in Region I}}\\
$^{209}$Bi$\to^{205}$Tl$$&3.138&${9/2^-}\to{1/2^+}$ &5&0.094&$6.34\times10^{26}$&{-}&{-}&{-}\\
$^{189}$Po$\to^{185}$Pb$$&7.694&${(5/2^-)}\to{3/2^-}$ &2&0.191&$3.80\times10^{-3}$&$1.25\times10^{-3}$&$6.53\times10^{-3}$&$7.14\times10^{-3}$\\
$^{203}$Po$\to^{199}$Pb$$&5.496&${5/2^-}\to{3/2^-}$ &2&0.134&$1.97\times10^{6}$&$8.15\times10^{5}$&$6.09\times10^{6}$&$5.99\times10^{6}$\\
$^{205}$Rn$\to^{201}$Po$$&6.386&${5/2^-}\to{3/2^-}$ &2&0.143&$6.88\times10^{2}$&$3.95\times10^{2}$&$2.77\times10^{3}$&$2.54\times10^{3}$\\
$^{207}$Ra$\to^{203}$Rn$$&7.269&${5/2^-\#}\to{3/2^-\#}$ &2&0.160&$1.60\times10^{0}$&$9.17\times10^{-1}$&$5.73\times10^{0}$&$5.24\times10^{0}$\\
$^{213}$Ra$\to^{209}$Rn$$&6.862&${1/2^-}\to{5/2^-}$ &2&0.126&$2.03\times10^{2}$&$2.61\times10^{1}$&$2.06\times10^{2}$&$2.08\times10^{2}$\\
$^{215}$Th$\to^{211}$Ra$$&7.665&${(1/2^-)}\to{5/2(^-)}$ &2&0.142&$1.20\times10^{0}$&$1.93\times10^{-1}$&$1.36\times10^{0}$&$1.51\times10^{0}$\\
$^{217}$U$\to^{213}$Th$$&8.425&${1/2^-\#}\to{5/2^-\#}$ &2&0.152&$8.00\times10^{-4}$&$4.11\times10^{-3}$&$2.70\times10^{-2}$&$3.14\times10^{-2}$\\
 \noalign{\global\arrayrulewidth1pt}\noalign{\global\arrayrulewidth0.4pt} \multicolumn{9}{c}{\textbf{Nuclei in Region II and III}}\\
$^{187}$Pb$\to^{183}$Hg$$&6.393&${3/2^-}\to{1/2^-}$ &2&0.166&$1.60\times10^{2}$&$8.89\times10^{0}$&$5.35\times10^{1}$&$9.39\times10^{1}$\\
$^{189}$Pb$\to^{185}$Hg$$&5.915&${3/2^-}\to{1/2^-}$ &2&0.163&$9.75\times10^{3}$&$1.03\times10^{3}$&$6.33\times10^{3}$&$1.09\times10^{4}$\\
$^{213}$Bi$\to^{209}$Tl$$&5.988&${9/2^-}\to{1/2^+}$ &5&0.092&$1.31\times10^{5}$&{-}&{-}&{-}\\
$^{223}$At$\to^{219}$Bi$$&4.723&${3/2^-\#}\to{9/2^-\#}$ &4&0.161&$6.25\times10^{5}$&{-}&{-}&{-}\\
$^{213}$Rn$\to^{209}$Po$$&8.245&${9/2^+\#}\to{1/2^-}$ &5&0.098&$1.95\times10^{-2}$&$8.47\times10^{-4}$&$8.65\times10^{-3}$&$8.09\times10^{-3}$\\
$^{219}$Rn$\to^{215}$Po$$&6.946&${5/2^+}\to{9/2^+}$ &2&0.193&$3.96\times10^{0}$&$1.02\times10^{0}$&$5.26\times10^{0}$&$6.19\times10^{0}$\\
$^{221}$Rn$\to^{217}$Po$$&6.162&${7/2^+}\to{(9/2^+)}$ &2&0.185&$6.98\times10^{3}$&$1.97\times10^{3}$&$1.07\times10^{4}$&$1.07\times10^{4}$\\
$^{221}$Fr$\to^{217}$At$$&6.457&${5/2^-}\to{9/2^-}$ &2&0.182&$2.88\times10^{2}$&$2.89\times10^{2}$&$1.59\times10^{3}$&$1.49\times10^{3}$\\
$^{215}$Ra$\to^{211}$Rn$$&8.864&${9/2^+\#}\to{1/2^-}$ &5&0.109&$1.67\times10^{-3}$&$8.07\times10^{-5}$&$7.39\times10^{-4}$&$7.36\times10^{-4}$\\
$^{219}$Ra$\to^{215}$Rn$$&8.138&${(7/2)^+}\to{9/2^+}$ &2&0.190&$1.00\times10^{-2}$&$5.88\times10^{-4}$&$3.08\times10^{-3}$&$3.47\times10^{-3}$\\
$^{217}$Th$\to^{213}$Ra$$&9.435&${9/2^+\#}\to{1/2^-}$ &5&0.122&$2.47\times10^{-4}$&$1.31\times10^{-5}$&$1.08\times10^{-4}$&$1.14\times10^{-4}$\\
$^{221}$Th$\to^{217}$Ra$$&8.625&${7/2^+\#}\to{(9/2^+)}$ &2&0.190&$1.78\times10^{-3}$&$1.23\times10^{-4}$&$6.47\times10^{-4}$&$6.35\times10^{-4}$\\

\end{tabular}
\end{ruledtabular}
\end{table*}

\begin{table*}
\caption{Same as Table \ref{table 1} and \ref{table 2}, but for favored $\mathcal{\alpha}$ decay of doubly-odd nuclei.}
\label{table 4}
\begin{ruledtabular}
\begin{tabular}{ccccccccc}
{$\mathcal{\alpha}$ transition} & $Q_{\alpha}$ (MeV) & ${j^{\pi}_{p}}\to{j^{\pi}_{d}}$ &$l_{\text{min}}$ &${P_{\alpha}}$ & $T^{\text{expt}}_{1/2}$ (s)&${T_{1/2}^{\text{calc1}}}$ (s)& ${T_{1/2}^{\text{calc2}}}$ (s)& ${T_{1/2}^{\text{calc3}}}$ (s)\\
\hline
 \noalign{\global\arrayrulewidth1pt}\noalign{\global\arrayrulewidth0.4pt} \multicolumn{9}{c}{\textbf{Nuclei in Region I}}\\
$^{192}$At$\to^{188}$Bi$$&7.696&${3^+\#}\to{3^+\#}$ &0&0.190&$1.15\times10^{-2}$&$1.44\times10^{-3}$&$7.60\times10^{-3}$&$8.41\times10^{-3}$\\
$^{200}$At$\to^{196}$Bi$$&6.596&${(3^+)}\to{(3^+)}$ &0&0.147&$8.26\times10^{1}$&$9.88\times10^{0}$&$6.73\times10^{1}$&$7.10\times10^{1}$\\
$^{202}$At$\to^{198}$Bi$$&6.353&${3(^+)}\to{3(^+)}$ &0&0.133&$1.45\times10^{3}$&$9.61\times10^{1}$&$7.23\times10^{2}$&$7.33\times10^{2}$\\
$^{204}$At$\to^{200}$Bi$$&6.071&${7^+}\to{7^+}$ &0&0.126&$1.43\times10^{4}$&$1.64\times10^{3}$&$1.30\times10^{4}$&$1.33\times10^{4}$\\
$^{206}$At$\to^{202}$Bi$$&5.886&${(5)^+}\to{5(^+\#)}$ &0&0.112&$2.02\times10^{5}$&$1.16\times10^{4}$&$1.03\times10^{5}$&$1.00\times10^{5}$\\
$^{208}$At$\to^{204}$Bi$$&5.751&${6^+}\to{6^+}$ &0&0.102&$1.06\times10^{6}$&$5.01\times10^{4}$&$4.90\times10^{5}$&$4.68\times10^{5}$\\
$^{200}$Fr$\to^{196}$At$$&7.615&${(3^+)}\to{(3^+)}$ &0&0.173&$4.75\times10^{-2}$&$1.36\times10^{-2}$&$7.86\times10^{-2}$&$7.47\times10^{-2}$\\
$^{204}$Fr$\to^{200}$At$$&7.170&${3^+}\to{(3^+)}$ &0&0.153&$1.82\times10^{0}$&$4.20\times10^{-1}$&$2.75\times10^{0}$&$2.71\times10^{0}$\\
$^{206}$Fr$\to^{202}$At$$&6.924&${3^+}\to{3(^+)}$ &0&0.139&$1.81\times10^{1}$&$3.31\times10^{0}$&$2.39\times10^{1}$&$2.33\times10^{1}$\\
$^{208}$Fr$\to^{204}$At$$&6.784&${7^+}\to{7^+}$ &0&0.129&$6.62\times10^{1}$&$1.09\times10^{1}$&$8.45\times10^{1}$&$8.49\times10^{1}$\\
$^{206}$Ac$\to^{202}$Fr$$&7.959&${(3^+)}\to{3^+}$ &0&0.172&$2.50\times10^{-2}$&$5.57\times10^{-3}$&$3.25\times10^{-2}$&$3.10\times10^{-2}$\\
 \noalign{\global\arrayrulewidth1pt}\noalign{\global\arrayrulewidth0.4pt} \multicolumn{9}{c}{\textbf{Nuclei in Region II and III}}\\
$^{214}$At$\to^{210}$Bi$$&8.987&${1^-}\to{1^-}$ &0&0.154&$5.58\times10^{-7}$&$1.05\times10^{-7}$&$6.80\times10^{-7}$&$7.02\times10^{-7}$\\
$^{216}$At$\to^{212}$Bi$$&7.950&${1(^-)}\to{1(^-)}$ &0&0.150&$3.00\times10^{-4}$&$7.42\times10^{-5}$&$4.95\times10^{-4}$&$4.76\times10^{-4}$\\
$^{218}$At$\to^{214}$Bi$$&6.874&${1^-\#}\to{1^-}$ &0&0.144&$1.50\times10^{0}$&$3.43\times10^{-1}$&$2.38\times10^{0}$&$2.11\times10^{0}$\\
$^{216}$Fr$\to^{212}$At$$&9.175&${(1^-)}\to{(1^-)}$ &0&0.161&$7.00\times10^{-7}$&$1.83\times10^{-7}$&$1.14\times10^{-6}$&$1.18\times10^{-6}$\\
$^{218}$Fr$\to^{214}$At$$&8.014&${1^-}\to{1^-}$ &0&0.166&$1.00\times10^{-3}$&$2.94\times10^{-4}$&$1.77\times10^{-3}$&$1.76\times10^{-3}$\\
$^{218}$Ac$\to^{214}$Fr$$&9.374&${1^-\#}\to{(1^-)}$ &0&0.169&$1.00\times10^{-6}$&$2.96\times10^{-7}$&$1.75\times10^{-6}$&$1.82\times10^{-6}$\\
$^{220}$Pa$\to^{216}$Ac$$&9.651&${1^-\#}\to{(1^-)}$ &0&0.178&$7.80\times10^{-7}$&$3.09\times10^{-7}$&$1.73\times10^{-6}$&$1.83\times10^{-6}$\\

\end{tabular}
\end{ruledtabular}
\end{table*}

\begin{table*}
\caption{Same as Table \ref{table 1} and \ref{table 2}, but for unfavored $\mathcal{\alpha}$ decay of doubly-odd nuclei.}
\label{table 5}
\begin{ruledtabular}
\begin{tabular}{ccccccccc}

{$\mathcal{\alpha}$ transition} & $Q_{\alpha}$ (MeV) & ${j^{\pi}_{p}}\to{j^{\pi}_{d}}$ &$l_{\text{min}}$ &${P_{\alpha}}$ & $T^{\text{expt}}_{1/2}$ (s)&${T_{1/2}^{\text{calc1}}}$ (s)& ${T_{1/2}^{\text{calc2}}}$ (s)& ${T_{1/2}^{\text{calc3}}}$ (s)\\
\hline
 \noalign{\global\arrayrulewidth1pt}\noalign{\global\arrayrulewidth0.4pt} \multicolumn{9}{c}{\textbf{Nuclei in Region I}}\\
$^{190}$Bi$\to^{186}$Tl$$&6.862&${(3^+)}\to{(2^-)}$ &1&0.163&$8.16\times10^{0}$&$2.01\times10^{-1}$&$1.23\times10^{0}$&$1.44\times10^{0}$\\
$^{192}$Bi$\to^{188}$Tl$$&6.381&${(3^+)}\to{(2^-)}$ &1&0.157&$2.77\times10^{2}$&$1.50\times10^{1}$&$9.54\times10^{1}$&$1.10\times10^{2}$\\
$^{194}$Bi$\to^{190}$Tl$$&5.918&${(3^+)}\to{2(^-)}$ &1&0.152&$2.05\times10^{4}$&$1.59\times10^{3}$&$1.04\times10^{4}$&$1.19\times10^{4}$\\
$^{210}$At$\to^{206}$Bi$$&5.631&${(5)^+}\to{6(^+)}$ &2&0.095&$1.66\times10^{7}$&$4.11\times10^{5}$&$4.34\times10^{6}$&$3.61\times10^{6}$\\
$^{210}$Fr$\to^{206}$At$$&6.672&${6^+}\to{(5)^+}$ &2&0.115&$2.67\times10^{2}$&$5.90\times10^{1}$&$5.11\times10^{2}$&$4.43\times10^{2}$\\
$^{212}$Fr$\to^{208}$At$$&6.529&${5^+}\to{6^+}$ &2&0.107&$2.78\times10^{3}$&$2.18\times10^{2}$&$2.04\times10^{3}$&$1.87\times10^{3}$\\
$^{212}$Pa$\to^{208}$Ac$$&8.415&${7^+\#}\to{(3^+)}$ &4&0.167&$7.50\times10^{-3}$&$1.07\times10^{-2}$&$6.43\times10^{-2}$&$5.89\times10^{-2}$\\
 \noalign{\global\arrayrulewidth1pt}\noalign{\global\arrayrulewidth0.4pt} \multicolumn{9}{c}{\textbf{Nuclei in Region II and III}}\\
$^{210}$Bi$\to^{206}$Tl$$&5.037&${1^-}\to{0^-}$ &2&0.082&$4.13\times10^{11}$&$6.60\times10^{7}$&$8.05\times10^{8}$&$8.54\times10^{8}$\\
$^{212}$Bi$\to^{208}$Tl$$&6.207&${1(^-)}\to{5^+}$ &5&0.079&$1.01\times10^{4}$&{-}&{-}&{-}\\
$^{214}$Bi$\to^{210}$Tl$$&5.621&${1^-}\to{5^+\#}$ &5&0.081&$5.66\times10^{6}$&{-}&{-}&{-}\\
$^{212}$At$\to^{208}$Bi$$&7.817&${(1^-)}\to{5^+}$ &5&0.077&$3.14\times10^{-1}$&$7.35\times10^{-3}$&$9.49\times10^{-2}$&$8.91\times10^{-2}$\\
$^{214}$Fr$\to^{210}$At$$&8.588&${(1^-)}\to{(5)^+}$ &5&0.089&$5.18\times10^{-3}$&$2.02\times10^{-4}$&$2.26\times10^{-3}$&$2.30\times10^{-3}$\\
$^{220}$Fr$\to^{216}$At$$&6.800&${1^+}\to{1(^-)}$ &1&0.163&$2.74\times10^{1}$&$6.73\times10^{0}$&$4.12\times10^{1}$&$4.02\times10^{1}$\\
$^{216}$Ac$\to^{212}$Fr$$&9.235&${(1^-)}\to{5^+}$ &5&0.103&$4.40\times10^{-4}$&$1.89\times10^{-5}$&$1.83\times10^{-4}$&$2.02\times10^{-4}$\\
$^{220}$Ac$\to^{216}$Fr$$&8.348&${(3^-)}\to{(1^-)}$ &2&0.171&$2.64\times10^{-2}$&$3.33\times10^{-4}$&$1.95\times10^{-3}$&$1.99\times10^{-3}$\\
\end{tabular}
\end{ruledtabular}
\end{table*}

\begin{table}[!htb]
\caption{The parameters of Eq. (\ref{subeq:21}) and that show $\mathcal{\alpha}$ preformation factors are linearly related to $N_pN_n$.}
\label{table 6}
\begin{ruledtabular}
\begin{tabular}{ccccc}

\multicolumn{1}{c}{\multirow{2}{*}{Region}}&\multicolumn{2}{c}{\text{favored decay}}&\multicolumn{2}{c}{\text{unfavored decay}}\\
 &a&b&a&b\\
\hline
\noalign{\global\arrayrulewidth1pt}\noalign{\global\arrayrulewidth0.4pt}&\multicolumn{4}{c}{\textbf{even-even Nuclei}}\\
I&-0.36222&0.14703&{-}&{-}\\
II, III&0.15948&0.21175&{-}&{-}\\
\noalign{\global\arrayrulewidth1pt}\noalign{\global\arrayrulewidth0.4pt} &\multicolumn{4}{c}{\textbf{odd-$A$ Nuclei}}\\
I&-0.34101&0.11712&-0.28777&0.11684\\
II, III&0.29582&0.16333&0.51621&0.09475\\
\noalign{\global\arrayrulewidth1pt}\noalign{\global\arrayrulewidth0.4pt}& \multicolumn{4}{c}{\textbf{doubly-odd Nuclei}}\\
I&-0.27858&0.09504&-0.33891&0.10868\\
II, III&0.22820&0.13944&0.55115&0.07457\\
\end{tabular}
\end{ruledtabular}
\end{table}

\begin{table}[!htb]
\caption{The standard deviations between $\mathcal{\alpha}$ decay half-lives of calculations and experimental data.}
\label{table 7}
\begin{ruledtabular}
\begin{tabular}{ccccccc}

\multicolumn{1}{c}{\multirow{2}{*}{Nuclei}}&\multicolumn{3}{c}{\text{favored decay}}&\multicolumn{3}{c}{\text{unfavored decay}}\\
 &$\sigma_1$&$\sigma_2$&$\sigma_3$&$\sigma_1$&$\sigma_2$&$\sigma_3$\\
\hline
even-even Nuclei&0.583&0.380&0.383\\
odd-$A$ Nuclei&0.659&0.370&0.366&0.897&0.542&0.536\\
doubly-odd Nuclei&0.813&0.215&0.213&1.631&0.940&0.926\\

\end{tabular}
\end{ruledtabular}
\end{table}

The aims of this work are to study the $\mathcal{\alpha}$ preformation factors and $\mathcal{\alpha}$ decay half-lives of nuclei around $Z=82$, $N=126$ shell closures. Many researchers suggested that the smaller valance nucleons (holes) nuclei have, the smaller $\mathcal{\alpha}$ preformation factors be \cite{PhysRevC.84.027303,Qian2013,PhysRevC.80.064325}. In 2011, Seif \textit{et al.} have put forward that the $P_{\alpha}$ of even-even nuclei around the $Z=82$, $N=126$ closed shells linearly depend on the product of the valance protons (holes) and neutrons (holes) $N_pN_n$ \cite{PhysRevC.84.064608}. Moreover, in our previous works, we systematically studied the $P_{\alpha}$ of the favored and unfavored $\mathcal{\alpha}$ decay for odd-$A$ and doubly-odd nuclei, which was extracted from the ratio of calculated $\mathcal{\alpha}$ decay half-life to the experimental data \cite{PhysRevC.94.024338,PhysRevC.96.024318}. The results indicated that the $P_{\alpha}$ is linearly related to the $N_pN_n$ although it is model-dependent. Recently, the CFM \cite{Rom.Rep.Phys123,0954-3899-40-6-065105,0954-3899-42-7-075106,PhysRevC.93.044326,SALEHAHMED2017103} was proposed to calculate the $P_{\alpha}$ with the difference of binding energy. It is a simple, effective and microscopic way. Once the binding energies of parent nuclei and neighboring nuclei are known, one can easily evaluate the $P_{\alpha}$. Therefore, it is interesting to validate whether the realistic $\mathcal{\alpha}$ preformation factor within CFM is also linearly dependent on $N_pN_n$. In addition, the Prox.1977 leaves $P_{\alpha}$ out of consideration or assumes as $P_{\alpha}=1$, thus the deviation between calculated $\mathcal{\alpha}$ decay half-life and experimental one is considerable \cite{PhysRevC.90.064603,Yao2015,PhysRevC.93.024612}. For confirming CFM and diminishing the difference between theoretical calculation and experimental data, in this work, we also calculate $\mathcal{\alpha}$ decay half-lives of 159 nuclei (including 50 even-even nuclei, 76 odd-$A$ nuclei and 33 doubly-odd nuclei) around $Z=82$, $N=126$ shell closures within Prox.1977 taking $P_{\alpha}=1$ and the realistic $P_{\alpha}$ evaluated by CFM, respectively.

For purpose of a simple description, we plot a nuclide distribution map in the Fig. \ref{fig 1}, and the area is divided into three regions by magic numbers ($Z=82$, $N=126$). In Region I, the proton numbers are above the $Z=82$ shell closure and the neutron numbers are below the $N=126$ closed shell, thus the $N_pN_n$ are negative. By that analogy, in Region II and III the $N_pN_n$ are positive. Therefore, both nuclei in Region II and III can be studied in an unified way. 

Firstly, we systematically calculate $\mathcal{\alpha}$ preformation factors within the CFM \cite{Rom.Rep.Phys123,0954-3899-40-6-065105,0954-3899-42-7-075106,PhysRevC.93.044326,SALEHAHMED2017103}. 
The results are listed in the fifth column of Table \ref{table 1}-\ref{table 5}. From these tables, we can find that the $P_{\alpha }$ sequence of nuclei from high to low is even-even nuclei, odd-$A$ nuclei and doubly-odd nuclei, which satisfy the variation tendencies of $P_{\alpha }$ obtained by various models \cite{ PhysRevC.45.2247,BUCK199353,PhysRevC.73.041301,PhysRevC.74.014304,PhysRevC.81.024315,PhysRevC.83.044317,PhysRevC.93.034316,PhysRevC.95.014319,PhysRevC.95.044303,1674-1137-41-1-014102}. In order to have a deeper insight into $P_{\alpha}$, we plot the relationship between $P_{\alpha}$ and $\frac{N_pN_n}{Z_0+N_0}$ of even-even nuclei, odd-$A$ nuclei (including favored and unfavored $\mathcal{\alpha}$ decay cases) and doubly-odd nuclei (including favored and unfavored $\mathcal{\alpha}$ decay cases) around $Z=82$, $N=126$ closed shells in Fig. \ref{fig 2}-\ref{fig 4}, respectively. In these figures, the red circle and blue triangle represent the cases of favored and unfavored $\mathcal{\alpha}$ decay, respectively. The red dash and blue solid lines represent the predictions of $\mathcal{\alpha}$ preformation factors for corresponding cases, which are expressed as
\begin{equation}
\
P_{\alpha}=a\frac{N_p N_n}{Z_0+N_0}+b
,\label{subeq:21}
\end {equation}
where $Z_0=82$ and $N_0=126$ represent the magic number of proton and neutron. The $a$ and $b$ are adjustable parameters, which are extracted from fittings of Fig. \ref{fig 2}-\ref{fig 4} and listed in Table \ref{table 6} (the left hand side for favored $\mathcal{\alpha}$ decaies and right hand side for unfavored ones). As shown in Fig. \ref{fig 2}-\ref{fig 4}, we can clearly see that all the $P_{\alpha}$ are linearly dependent on $N_pN_n$ for cases of even-even nuclei, odd-$A$ nuclei and doubly-odd nuclei. It indicates that valance proton-neutron interaction plays a key role in the $\mathcal{\alpha}$ preformation and the influence of proton-neutron pairs on the $\mathcal{\alpha}$ cluster basically maintain invariable in the same region. In the Fig. \ref{fig 3}, we can distinctly find that the linear relationship between $P_{\alpha}$ and $N_pN_n$ for the cases of even-odd and odd-even nuclei without obvious difference. It manifests that in the $N_pN_n$ scheme, the effect of unpaired odd neutron or proton on $P_{\alpha}$ can be treated in an unified way. It also verifies that using different methods to calculate $P_{\alpha}$ of even-odd nuclei and odd-even nuclei in the CFM is appropriate. Combining with our previous works \cite{PhysRevC.94.024338,PhysRevC.96.024318}, we confirm that the $P_{\alpha}$ of nuclei around $Z=82$, $N=126$ closed shells is linearly dependent on $N_pN_n$ whether the $P_{\alpha}$ is model-dependent or microcosmic.  

Secondly, we systematically calculate $\mathcal{\alpha}$ decay half-lives of these nuclei within Prox.1977. The experimental $\mathcal{\alpha}$ decay half-lives are taken from the latest evaluated nuclear properties table NUBASE2016 \cite{1674-1137-41-3-030001}, the $\mathcal{\alpha}$ decay energies are taken from the latest evaluated atomic mass table AME2016 \cite{1674-1137-41-3-030002,1674-1137-41-3-030003}. The detailed calculations are listed in Table \ref{table 1}-\ref{table 5}. In these tables, the first four columns denote $\mathcal{\alpha}$ decay, experimental decay energy, spin and parity transition and the minimum angular momentum taken away by the $\mathcal{\alpha}$ particle, respectively. The fifth one is $\mathcal{\alpha}$ preformation factors calculated with CFM. The sixth one denotes experimental $\mathcal{\alpha}$ decay half-life. The last three ones are calculated $\mathcal{\alpha}$ decay half-life by Prox.1977 without considering $P_{\alpha}$, with taking $P_{\alpha}$ by CFM and with fitting $P_{\alpha}$ calculated by Eq. (\ref{subeq:21}) and parameters listed in Table \ref{table 6}, which are denoted as ${T_{1/2}^{\text{calc1}}}$, ${T_{1/2}^{\text{calc2}}}$ and ${T_{1/2}^{\text{calc3}}}$, respectively. All tables are divided into two parts: the upper half part is nuclei in Region I and the lower one is nuclei in Region II and III. From Table \ref{table 1}-\ref{table 5} we find that although the $T_{1/2}^{\text{calc1}}$ can produce experimental data, the deviation is still considerable. So we calculate decay constant $\lambda$ with $P_{\alpha}$, which is evaluated by CFM. The new calculated $\mathcal{\alpha}$ decay half-lives $T_{1/2}^{\text{calc2}}$ can better reproduce with $T_{1/2}^{\text{expt}}$ than $T_{1/2}^{\text{calc1}}$. In addition, we can find the $T_{1/2}^{\text{calc3}}$, which is calculated with fitting $P_{\alpha}$, can well conform the $T_{1/2}^{\text{calc2}}$. It indicates that $P_{\alpha}$ is linearly related to $N_pN_n$ well. In order to intuitively survey the deviations between $\mathcal{\alpha}$ decay half-lives of calculations and experimental data, we calculate the standard deviation $\sigma=\sqrt{\sum ({\log_{10}T^{\rm{calc}}_{1/2}-\log_{10}T^{\rm{expt}}_{1/2}})^2/n}$. The results $\sigma_1$, $\sigma_2$ and $\sigma_3$ denote standard deviations between $T_{1/2}^{\text{calc1}}$, $T_{1/2}^{\text{calc2}}$, $T_{1/2}^{\text{calc2}}$ and $T_{1/2}^{\text{expt}}$, respectively, which are listed in Table \ref{table 7}. In this table, we can clearly see that the values of $\sigma_2$ significantly reduce compared to $\sigma_1$ and the $\sigma_2$ are basically equal to $\sigma_3$.                                                                                                                                                                                                                                                                                                                                                                                                                                                                                                                                                                                                                                                                                                                                                                                  It indicates that the calculations within Prox.1977 using $P_{\alpha}$ from CFM can better reproduce with experimental data than using $P_{\alpha}=1$ as well as the $P_{\alpha}$ have  linear relationship with $N_pN_n$. For nuclei $^{209}$Bi, $^{213}$Bi and $^{223}$At in Table \ref{table 3} as well as nuclei $^{212}$Bi and $^{214}$Bi in Table \ref{table 5}, we cannot obtain the classical turning points $r_{\text{in}}$ through solving equation $V(r_{\text{in}})=V(r_{\text{out}})=Q_{\alpha}$ due to the depths of potential well above the $Q_{\alpha}$. Therefore, we don\rq{}t give the calculations of half-lives for these 5 nuclei. This phenomenon motivate our interesting to further develop the theoretical model in the future.
\section{Summary}
\label{section 4}

In summary, we preformed the systematically study of $\mathcal{\alpha}$ preformation factors within the cluster-formation model (CFM) and $\mathcal{\alpha}$ decay half-lives within the proximity potential 1977 formalism (Prox.1977) for nuclei around $Z=82$, $N=126$ closed shells. Our results indicate that the realistic $P_{\alpha}$ calculated by CFM for nuclei around $Z=82$, $N=126$ shell closures are linear with $N_pN_n$. Combining with our previous works, it confirms that valance proton-neutron plays an important role in the $\mathcal{\alpha}$ cluster formation. In addition, our calculated $\mathcal{\alpha}$ decay half-lives i.e. $T_{1/2}^{\text{calc2}}$, using Prox.1977 taking $P_{\alpha}$ evaluated by CFM, can well reproduce the experimental data and significantly reduce the errors. It demonstrates that the CFM is credible. This work will be a reference for future experiments and theoretical researches.

\begin{acknowledgments}

We would like to thank Xiao-Dong Sun, Hao-Jie Shen, Tian Huang and Xin Li for useful discussion. This work is supported in part by the National Natural Science Foundation of China (Grants No. 11205083 and No. 11505100), the Construct Program of the Key Discipline in Hunan Province, the Research Foundation of Education Bureau of Hunan Province, China (Grant No. 15A159), the Natural Science Foundation of Hunan Province, China (Grants No. 2015JJ3103 and No. 2015JJ2121), the Innovation Group of Nuclear and Particle Physics in USC, the Shandong Province Natural Science Foundation, China (Grant No. ZR2015AQ007), Hunan Provincial Innovation Foundation For Postgraduate (Grant No. CX2017B536).

\end{acknowledgments}

%\bibliographystyle{apsrev4-1}
%\bibliography{reference}


\end{document}